\documentclass[aps,prb,showpacs,superscriptaddress,twocolumn,10pt,floatfix]{revtex4-1}
\usepackage{graphicx}
\usepackage{natbib}
\usepackage{amsmath}
\usepackage{epsfig}
\usepackage{physics}
\usepackage{multirow}
\usepackage{siunitx}
\usepackage{natbib}
\usepackage{indentfirst}
\usepackage{siunitx}
\usepackage{mathtools}
\usepackage{amssymb}
\usepackage{bbold}
\usepackage{xcolor}
\usepackage{bm}
\usepackage{eufrak}
\usepackage{epstopdf}
\usepackage{fancyhdr}
\usepackage{ulem}
\begin{document}
	
	\title{Beating Fabry-P\'erot interference pattern in a magnonic scattering junction in the graphene quantum Hall ferromagnet}
	
	\author{Jonathan Atteia}
	\email{jonathan.atteia@cnrs.fr}
	\affiliation{Laboratoire de Physique des Solides, Universit\'e Paris Saclay, CNRS UMR 8502, F-91405 Orsay Cedex, France}
	\affiliation{Aix-Marseille Universit\'e, CNRS, CINaM, Marseille, France}
	
	\author{Preden Roulleau}
	\affiliation{SPEC, CEA, CNRS, Universit\'e Paris-Saclay, CEA Saclay, F-91191 Gif sur Yvette Cedex, France}
	
	\author{Mark O. Goerbig}
	\email{mark-oliver.goerbig@universite-paris-saclay.fr}
	\affiliation{Laboratoire de Physique des Solides, Universit\'e Paris Saclay, CNRS UMR 8502, F-91405 Orsay Cedex, France}
	
	\date{\today}
	\begin{abstract}
		
		At filling factor $\nu=0,\pm1$, the ground state of graphene is a particular SU(4) ferromagnet which hosts a rich phase diagram along with several spin, pseudospin or ``entanglement'' magnon modes. Motivated by recent experiments, we study a $\nu=-1|0|-1$ Fabry-Pérot magnonic junction. If the ground state at $\nu=0$ is spin polarized, there exist two spin modes which interfere and create a beating pattern, while pseudospin modes are reflected. The same scenario occurs for pseudospin magnons if the $\nu=0$ ground state is spin polarized. The observation of such an interference pattern would provide information on the low-energy anisotropies and thus on the ground state.
		
	\end{abstract}
	
	\maketitle

\section{Introduction}

Magnons, the quanta of spin waves, are promising candidates for the transport of information in quantum devices based on ferromagnetic and antiferromagnetic materials\cite{Wolf2001,Chumak2015}. In traditional ferromagnets, a magnon corresponds simply to a spin excitation propagating across the sample without any electric current, allowing to transfer information without dissipation. A special type of two-dimensional (2D) ferromagnetism arises in 2D electron systems exposed to a strong perpendicular magnetic field that quantizes the electronic bands into highly degenerate Landau levels (LLs). It consists of a spontaneous spin polarization to minimize the Coulomb interaction. A particular example of such systems is graphene, where in addition to the electron spin, the valley degree of freedom yields a pseudospin, which leads to a particular SU(4) ferromagnetism\cite{Nomura2006,Young2012,Goerbig2011}. 
At charge neutrality, a rich phase diagram has been presented with different phases characterized by their spin and valley polarizations\cite{Kharitonov2012}. Inherited from these phases, graphene also possesses a very rich magnonic structure\cite{Atteia2021b} where, along with the usual spin magnon, there also exist pseudospin magnons as well as more exotic \textit{entanglement}\cite{Doucot2008} magnons where both spin and pseudospin are flipped. 

Due to the gate tunability of the filling factor and the ability to realize high-quality samples, graphene is thus a promising platform for the study of SU(4) spintronics and magnonics. This field has emerged recently with the experimental emission and detection of magnons at filling factor $\nu=\pm1$ using local gates tuned at  $\nu=2$\cite{Stepanov2018a,Wei2018,Assouline2021,Fu2021a,Pierce2022}. Theoretically, it led to various proposals aimed at investigating the physical properties of magnons in graphene. For instance, it has been shown in Ref. [\onlinecite{Wei2021}] that the electrical dipole carried by magnons can interact with the electric field formed at a $\nu=-1|\nu$=1 junction. To ensure the success of further experimental magnonic experiments, it is crucial to investigate the following fundamental properties. (\textit{i}) Different types of magnons can coexist : spin magnons or pseudospin magnons. From an experimental perspective, it is crucial to develop methods that allow for the selective emission of specific types of magnons. (\textit{ii}) 
The physical mechanism underlying the properties of magnons relies on the sublattice potential $\Delta_{AB}$, the Zeeman effect $\Delta_{Z}$ and different anisotropy factors. To date, there are no experimental estimates for these terms. Our original magnon setup is specifically designed to address these fundamental properties. 

\begin{figure}[h!]
	\begin{center}
		\includegraphics[width=7cm]{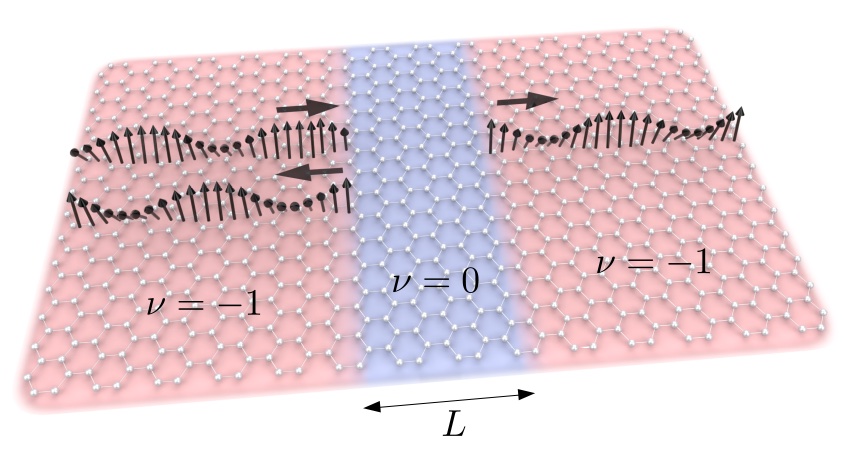} 
		\caption{a) Scattering setup. A magnon is injected in the left region ($\nu=-1$), it is scattered by a region of length $L$ at $\nu=0$, while its transmission is measured in the right region ($\nu=-1$).}
		\label{fig:scat_set}
	\end{center}
\end{figure}

\begin{figure*}
	\begin{center}
		a)\includegraphics[width=0.95\textwidth]{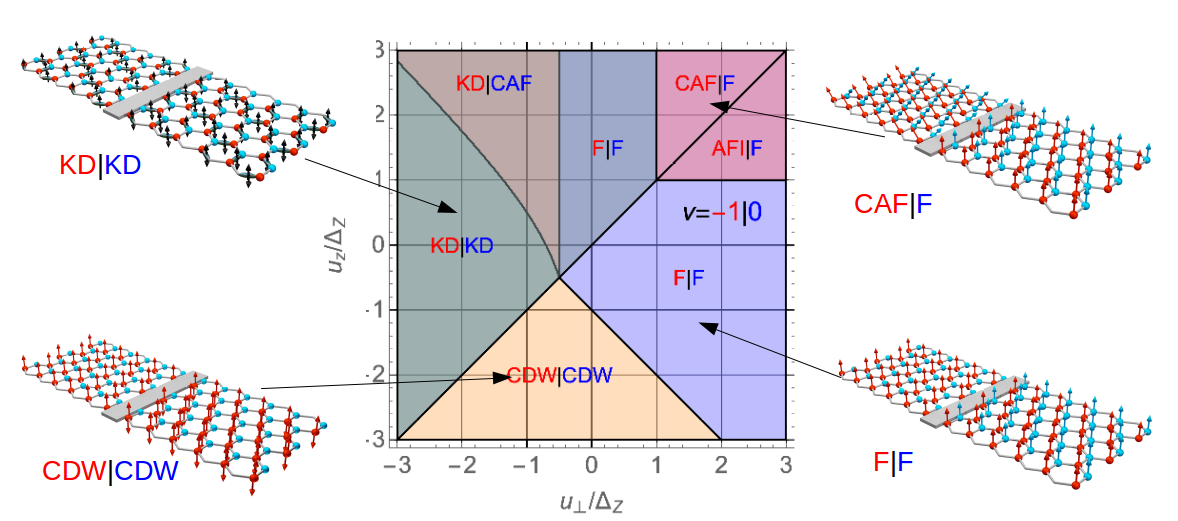} 
		b)\includegraphics[width=0.44\textwidth]{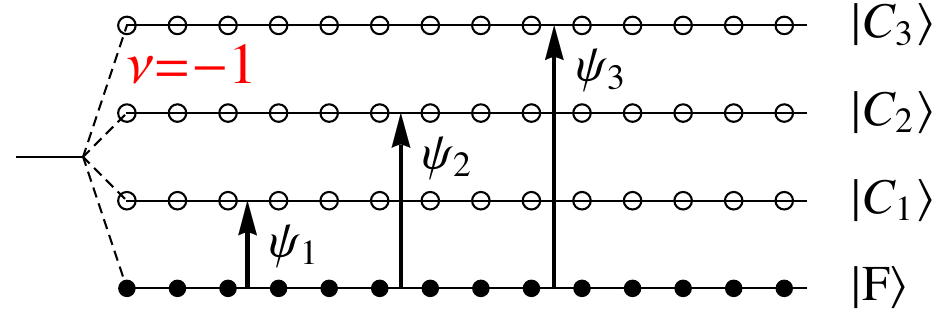}
		c)\includegraphics[width=0.44\textwidth]{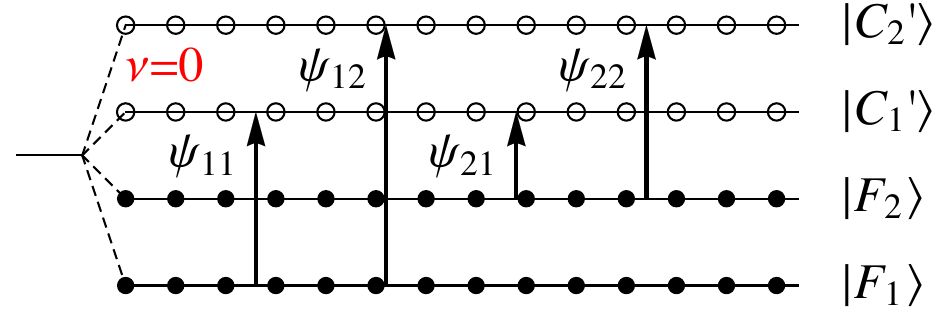}
		\caption{a) Superposition of the $\nu=-1$ and $\nu=0$ phase diagrams, as a function of the free parameters $u_\perp/\Delta_Z$ and $u_z/\Delta_Z$. The labels $A|B$ correspond to the phase $A$ at $\nu=-1$ and the phase $B$ at $\nu=0$. The insets show sketches of the spin and electron configurations on the different sublattices across the interface for some of the phases discussed in the main text. Sub-LL magnon transitions at b) $\nu=-1$ and c) $\nu=0$. At $\nu=-1$, one sub-LL is filled in the state $|F\rangle$ and there are three magnon modes described by the fields $\psi_i$. At $\nu=0$, two sub-LL are filled and there are four magnon modes $\psi_{ab}$ going from the sub-LL $|F_a\rangle$ to $|C_b'\rangle$.}
		\label{fig:LLSW}
	\end{center}
\end{figure*}

\section{Model and phase diagram}

We consider a $\nu=-1|0|-1$ magnonic Fabry-P\'erot interference setup as shown in Fig. \ref{fig:scat_set}. A magnon is injected at $\nu=-1$ in the left region, it is scattered at the interfaces with the $\nu=0$ central region of length $L$, and the transmission probability $T$ is measured in the $\nu=-1$ right region. Due to particle-hole symmetry, the setup is equivalent to $\nu=1|0|1$, and our choice is simply guided by illustration purposes. We investigate the magnon transmission as a function of the phases formed in the two types of region [see phase diagram of Fig. \ref{fig:LLSW}.a)]. 

At filling factor $\nu=\pm1$, the SU(4) ferromagnet is composed of four phases: two ferromagnetic and two entangled ones\cite{Lian2016,Lian2017,SuppMat}. For simplicity, we consider only the two spin ferromagnetic phases with different accompanying pseudospin polarization (Fig. \ref{fig:LLSW}). In the ferromagnetic charge density wave phase (F-CDW) phase, the electrons are polarized in one valley and thus on a single sublattice ($|F\rangle=|K\uparrow\rangle$) whereas in the ferromagnetic Kekul\'e-distorted (F-KD) phase, the pseudospin is in a coherent superposition of the two valleys [$|F\rangle=\frac{1}{\sqrt{2}}(|K\rangle+e^{i\phi}|K'\rangle)|\uparrow\rangle$]. Since the $\nu=\pm1$ phases are spin \textit{and} pseudospin polarized, when matching the $\nu=-1$ region with $\nu=0$, we shall refer to the $\nu=\pm1$ phases either by their spin polarization (F) or their pseudospin polarization (CDW or KD). At $\nu=0$ (charge neutrality), there are also four phases\cite{Kharitonov2012,SuppMat} which have all been observed experimentally in different setups : (i) two fully or partially spin polarized phases, the ferromagnetic\cite{Veyrat2020b} (F) and canted antiferromagnetic\cite{Young2014} (CAF) phases, and (ii) two pseudospin polarized phases, the CDW \cite{Coissard2022}  and the KD \cite{Li2019,Liu2022} phase. As mentioned above, which of these phases is encountered is highly sample-dependent. From a theoretical point of view, one expects the SU(4)-ferromagnetic phase to be fixed by subleading energy scales below the SU(4)-invariant Coulomb interaction. The first one to mention is the Zeeman effect $\Delta_Z$ which favors a spin polarization in the direction of the magnetic field. Furthermore, an underlying boron-nitride substrate may induce a sublattice potential $\Delta_{AB}$ in the graphene sheet that acts as a Zeeman effect on the valley pseudospin. Finally, one needs to take into account local pseudospin anisotropies $u_z$ and $u_\perp$ originating from shortrange electron-electron interactions, or electron-phonon coupling\cite{Alicea2006,Nomura2009,Kharitonov2012}. We will show in the following that  the observable beating patterns and threshold energies observed in our magnonic Fabry-Pérot may provide experimental insight into their values. 

Let us consider the setup shown in Fig. \ref{fig:scat_set} with a spin-polarized  ground state in the central $\nu=0$ region. There exist two spin and two entanglement modes such that an incoming spin magnon at $\nu=-1$ can propagate into the $\nu=0$ region while a pseudospin magnon will be fully reflected. On the other hand, if the ground state is pseudospin polarized, there are two pseudospin modes such that pseudospin magnons can propagate while spin magnons are totally reflected. 
These properties can easily be studied experimentally, and we will concentrate on these two configurations in the following. To appreciate this effect, we need to consider the different transitions between phase states as depicted in Fig. 2.b) and c). We begin by considering the spin polarized F-CDW phase at $\nu=-1$ and the $\nu=0$ F phase. At $\nu=-1$, we have $|F\rangle=|K\uparrow\rangle$, while at $\nu=0$, we have $|F_1\rangle=|F\rangle$ and $|F_2\rangle=|K'\uparrow\rangle=|C_1\rangle$. The field $\psi_1$ at $\nu=-1$, which corresponds to a $|K\uparrow\rangle\rightarrow|K'\uparrow\rangle$ transition namely a pseudospin magnon is thus blocked at $\nu=0$ because the sub-LL $|F_2\rangle$ is filled. On the other hand, the field $\psi_2$, which corresponds to a pure spin magnon sketched in Fig. \ref{fig:LLSW}.b), matches $\psi_{11}$ which, along with $\psi_{22}$, is also a spin transition [see Fig. \ref{fig:LLSW}.c)]. However, as discussed below, the magnonic eigenstates are not necessarily the $\psi_{ab}$ fields but rather superpositions such that the $\psi_2$ fields at $\nu=-1$ may excite more than one eigenmode at $\nu=0$.

We will demonstrate that this property proves to be highly insightful in extracting $\Delta_{AB}$, $\Delta_{Z}$ and different anisotropy factors. Indeed, in the $\nu$ = 0 region, there exist two distinct magnon modes, either with two spin magnons or two pseudospin magnons, characterized by gaps $\Delta^1$ and $\Delta^2$ and possessing different wavevectors. As a result, the two magnonic channels interfere, giving rise to a beating pattern of the transmission.  The observation of such an interference pattern would allow for an experimental access to the values of the spin or pseudospin gaps $\Delta^1$ and $\Delta^2$ and therefore  would give valuable information about the values of valley anisotropies and consequently about the location of the sample in the phase diagram (see Table I).

\section{Magnons dispersion relations}

Our starting point is the non-linear sigma model widely used in this domain\cite{Arovas1999,Yang2006,Nomura2009,Lian2017,Atteia2021a,Atteia2021b,SuppMat} (See Appendix \ref{app:nlsm} for a short review). At $\nu=-1$, to introduce a magnon, we perform small periodic unitary SU(4) rotations of the state $|F\rangle$ to the state $|C_i\rangle$ parametrized by three complex fields $\psi_i(\mathbf{r},t)$ with $i\in\{1,2,3\}$[see Fig. \ref{fig:LLSW}.b)], such that we define the field $Z(\mathbf{r},t)$
\begin{align}
	Z(\mathbf{r},t)&=e^{i(\psi_i(\mathbf{r},t)\Gamma_i+\psi_i(\mathbf{r},t)^*\Gamma^\dagger_i)}|F\rangle \\
	&=\left[1+i(\psi_i(\mathbf{r},t)\Gamma_i+\psi_i^*(\mathbf{r},t)\Gamma^\dagger_i)-\frac{1}{2}|\psi_i|^2\right]|F\rangle 
\end{align}
where summation over repeated indices is assumed, and the $\Gamma_i$ matrices are the generalization of the spin-flip Pauli matrices $\sigma_\pm=\sigma_x\pm i\sigma_y$ to the SU(4) formalism. In the second line, we have assumed small perturbations around the ground state $|\psi_i|\ll1$ which implies low-energy magnons.  The generators for a ground state $|F\rangle=(1,0,0,0)^T$ can be expressed in a compact way as
\begin{align}
	\psi_i\Gamma_i+\psi_i^*\Gamma^\dagger_i=
	\begin{pmatrix}
		0 & \psi_1^* &\psi_2^*&\psi_3^* \\\psi_1&0&0&0 \\ \psi_2& 0 &0&0 \\\psi_3& 0 &0&0 
	\end{pmatrix}
\end{align}
We can see that the fields $\psi_i$ generate a transition from the ground state $|F\rangle$ to the state $|C_i\rangle$, whereas the $\psi^*_i$ are the particle-hole conjugate partners which generate transitions from the states $|C_i\rangle$ to $|F\rangle$. The $\psi^*_i$ fields are thus the particle-hole conjugate of the $\psi_i$ fields. The particle-hole conjugate of the $\nu=-1$ ground state is the $\nu=1$ ground state and the $\psi^*_i$ fields act thus on the $|F'\rangle$ spinor at $\nu=1$. These spinors correspond thus to antimagnons.

Analogously to the $\nu=\pm1$ case, at $\nu=0$, we perform similar unitary rotations of the ground state on the two spinors
\begin{align}
	Z_1(\mathbf{r},t)=e^{i(\psi_{1a}\Gamma_{1a}+\psi_{1a}^*\Gamma_{1a}^\dagger)}|F_1\rangle \\
	Z_2(\mathbf{r},t)=e^{i(\psi_{2a}\Gamma_{2a}+\psi_{2a}^*\Gamma_{2a}^\dagger)}|F_2\rangle
\end{align}
where the $\Gamma_{ab}$ matrices can be defined as
\begin{align}
	\psi_{ab}\Gamma_{ab}+\psi_{ab}^*\Gamma_{ab}^\dagger=\begin{pmatrix}
		0 & 0 &\psi_{11}^*&\psi_{12}^* \\0&0&\psi_{21}^*&\psi_{22}^*\\ \psi_{11}& \psi_{21}&0&0 \\\psi_{12}& \psi_{22}&0&0 
	\end{pmatrix}
\end{align}
We can see that the matrix $\Gamma_{ab}$ generates a flip from the level $|F_a\rangle$ to the level $|C'_b\rangle$ parametrized by the four complex fields $\psi_{ab}$ [see Fig. \ref{fig:LLSW}.c)] with $a,b\in\{1,2\}$ and $|\psi_{ab}|\ll1$. 

Expanding the non-linear sigma model Lagrangian (See Appendix \ref{eq:NLSM}) up to second order in the $\psi$ fields ($\mathcal{L}\approx\mathcal{L}^{(0)}+\mathcal{L}^{(2)}$) along the lines of Ref. [\onlinecite{Atteia2021b}]. Minimizing the action, we obtain a generalized Bogolioubov-de Gennes-Schrödinger equation of motion for the magnons
\begin{align}
	i\tau_z\partial_t\Psi_\nu(\mathbf{r},t)=-\frac{\rho_s}{n_0/2}\bm{\nabla}^2\Psi_\nu(\mathbf{r},t)-R_\nu[P_0]\Psi_\nu(\mathbf{r},t),
	\label{eq:Schrodinger}
\end{align}
where $\Psi_{\nu=-1}=(\psi_1,\psi_2,\psi_3,\psi_1^*,\psi_2^*,\psi_3^*)$ is the spinor field at $\nu=-1$ which encompasses the three magnon modes as well as their complex conjugate fields $\psi_i^*$. Similarly,  $\Psi_{\nu=0}=(\psi_{11},\psi_{12},\psi_{21},\psi_{22},\psi_{11}^*,\psi_{12}^*,\psi_{21}^*,\psi_{22}^*)$ encodes the four magnon and antimagnon modes at $\nu=0$,  $\rho_s=(1/16\sqrt{2\pi})e^2/\varepsilon l_B$ is the spin stiffness with the magnetic length $l_B=\sqrt{\hbar/eB}$. Moreover, $n_0=(2\pi l_B^2)^{-1}$ is the electronic density in a sub-LL and $\tau_z\equiv\sigma_z\otimes\mathbb{1}_{3\times 3}$ at $\nu=-1$ or $\tau_z\equiv\sigma_z\otimes\mathbb{1}_{4\times 4}$ at $\nu=0$, where $\sigma_z$ is the Pauli matrix acting in magnon/antimagnon space. The anisotropic matrix $R_\nu[P_0]$ depends on the ground state order parameter $P_0$ at filling factor $\nu$ and encodes all information about the gaps and the coupling between the modes at $\nu=-1$ and $\nu=0$. It can be expressed as
\begin{align}
	R_\nu[P_0]=\begin{pmatrix}
		M & N^\dagger \\ N & M ^T
	\end{pmatrix}
\end{align}
where $N$ and $M$ are $3\times3$ matrices at $\nu=-1$ and $4\times4$ matrices at $\nu=-1$. In the CDW and F phases, the matrix $N$ vanishes and $R_\nu[P_0]$ is thus block-diagonal such that the positive and negative energy sectors are decoupled. We can thus consider only the $\psi$ fields and neglect the $\psi^*$ fields. On the contrary, in the KD and CAF phase, the negative and positive energy sectors are coupled. This coupling is responsible for linearly dispersing magnon modes associated with a U(1) symmetry of the ground state\cite{Atteia2021b,Nomura2009} analogously to superfluidity. 

Due to invariance under space and time translations, we first express the spinor field $\Psi(\mathbf{r},t)$ as a plane wave $\Psi(\mathbf{r},t)=\Psi e^{i(\mathbf{k}\cdot\mathbf{r}-E t)}$ and diagonalize Eq. (\ref{eq:Schrodinger}). For illustrative purposes, we concentrate on the CDW and F phases at $\nu=0$, in which the magnon and antimagnon subspaces are decoupled. We therefore need to focus only on the spinors $\Psi^\alpha_{\nu}$ for the eigenmode $\alpha$ at filling factor $\nu$ associated with the positive energies. In these phases, the dispersion of the mode $\alpha$ is quadratic
\begin{align}
	E_\nu^\alpha(\mathbf{k})=4\pi\rho_s(\mathbf{k}l_B)^2+\Delta_\nu^\alpha
	\label{eq:disp}
\end{align}
where the first term is the SU(4)-invariant part and $\Delta_\nu^\alpha$ is the gap associated with the mode $\alpha$ at filling factor $\nu$. Table \ref{tab:gaps} presents the explicit expression for the gaps of the modes $\alpha=1$ and $\alpha=2$ at $\nu=0$ which correspond to the two spin modes in the spin-polarized phases or the two pseudospin modes in the pseudospin-polarized phases. 

As mentioned above, we focus on the CDW and F phases but the results remain the same for other phases as long as the $\nu=0$ and $\nu=-1$ regions are in the same phase. In the CDW and F phases at $\nu=0$, the spinors of the four modes are given by
\begin{align}
	\Psi^1_{\nu=0}=\phi_0\begin{pmatrix} -1 \\ 0 \\ 0 \\ 1\end{pmatrix} \quad
	\Psi^2_{\nu=0}=\phi_0\begin{pmatrix} 1 \\ 0 \\ 0 \\ 1\end{pmatrix} \\
	\Psi^3_{\nu=0}=\phi_0\begin{pmatrix} 0 \\ 1 \\ 0 \\ 0\end{pmatrix} \quad
	\Psi^4_{\nu=0}=\phi_0\begin{pmatrix} 0 \\ 0 \\ 1 \\ 0\end{pmatrix} 
\end{align}
where $\Psi_{\nu=0}^\alpha=(\psi_{11},\psi_{12},\psi_{21},\psi_{22})$ and $\phi_0\ll 1$ is the amplitude of the spin wave. We can see that the eigenmodes $\alpha=1$ and $\alpha=2$ are superpositions of the spin waves transition $\psi_{11}$ and $\psi_{22}$. In the F phase, according to our convention, $\psi_{11}$ is associated with the transition $|K\uparrow\rangle\rightarrow|K\downarrow\rangle$ whereas $\psi_{22}$ is associated with the transition $|K'\uparrow\rangle\rightarrow|K'\downarrow\rangle$ such that both transitions are of spin type. On the other hand, in the (pseudospin polarized) CDW phase, the transitions $\psi_{11}$ and $\psi_{22}$ refer to pseudospin modes. Finally, $\psi_{12}$ and $\psi_{21}$ represent entanglement transitions where both spin and pseudospin are simultaneously flipped. These modes are decoupled from the others and not considered in our scattering setup.

\section{Scattering at the interfaces}

Let us now describe theoretical magnon transport through the setup shown in Fig. \ref{fig:scat_set}. At $\nu=-1$, we inject either a spin magnon ($i=1$) described by the field $\psi_1$ or a pseudospin magnon ($i=2$) described by the field $\psi_2$. We consider the two cases, (i) an incident \textit{spin} magnon, scattered by a \textit{spin polarized} $\nu=0$ region or (ii) and incident \textit{pseudospin} magnon scattered by a \textit{pseudospin polarized} phase, and calculate the transmission probability of the magnon in the right region. We consider periodic boundary conditions along the $y$ direction such that $k_y$ is a good quantum number. We construct the scattering states as a superposition of the eigenmodes in the tree regions, namely
\begin{align}
	\Psi^L(\mathbf{r},t)&=\left(\Psi_{\nu=-1}^i e^{ik_{x,-1}^i x}+\sum_{\alpha=1}^3r_\alpha\Psi_{\nu=-1}^\alpha e^{-ik_{x,-1}^\alpha x}\right)e^{ik_yy}\\
	\Psi^C(\mathbf{r},t)&=\left(\sum_{\alpha=1}^4c_\alpha^+\Psi_{\nu=0}^\alpha e^{ik_{x,0}^\alpha x}+c_\alpha^-\Psi_{\nu=0}^\alpha e^{-ik_{x,0}^\alpha x}\right)e^{ik_yy} \\
	\Psi^R(\mathbf{r},t)&=\left(\sum_{\alpha=1}^3t_\alpha\Psi_{\nu=-1}^\alpha e^{ik_{x,-1}^\alpha x}\right)e^{ik_yy}
\end{align}
where $\Psi_{\nu=-1}^\alpha=(\psi_1,\psi_2,\psi_3)$ are the three eigenmodes at $\nu=-1$, $r_\alpha$ and $t_\alpha$ are the reflection and transmission coefficients in the mode $\alpha$ and $i\in\{1,2\}$ indicates an incident spin or pseudospin magnon. The wavevectors $k_{x,\nu}^\alpha\equiv k_{x,\nu}^\alpha(E)$ are obtained by inverting the dispersion relation (\ref{eq:disp}) for the mode $\alpha$ at filling factor $\nu$. The weights $c_\alpha^+$ and $c_\alpha^-$ of the wavefunctions in mode $\alpha$ in the central region are not of interest here.

\begin{figure*}
	a)\includegraphics[width=0.32\textwidth]{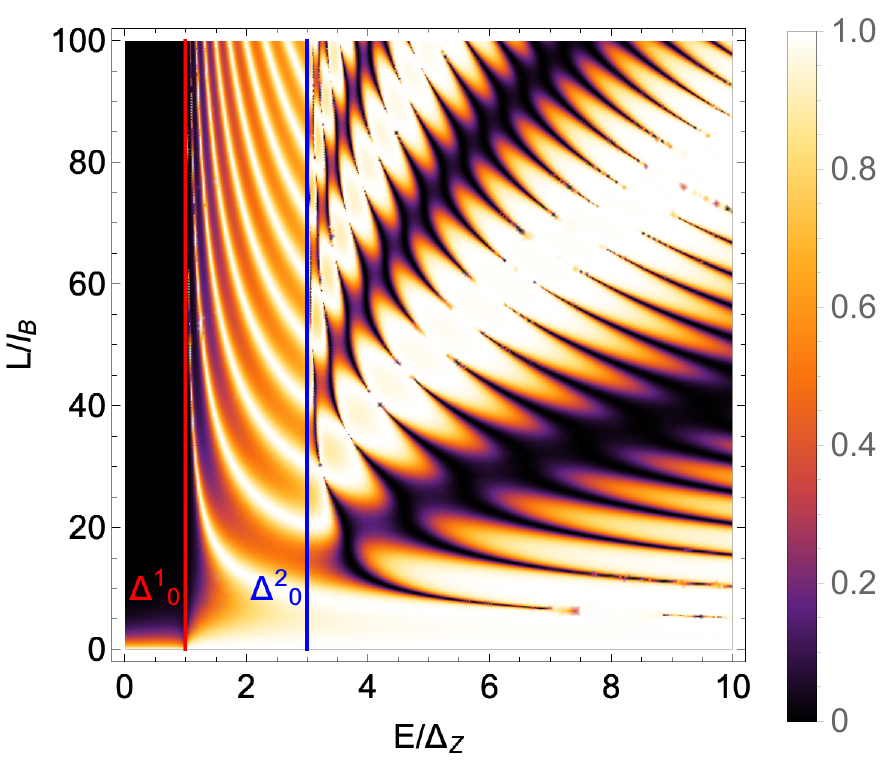}
	b)\includegraphics[width=0.29\textwidth]{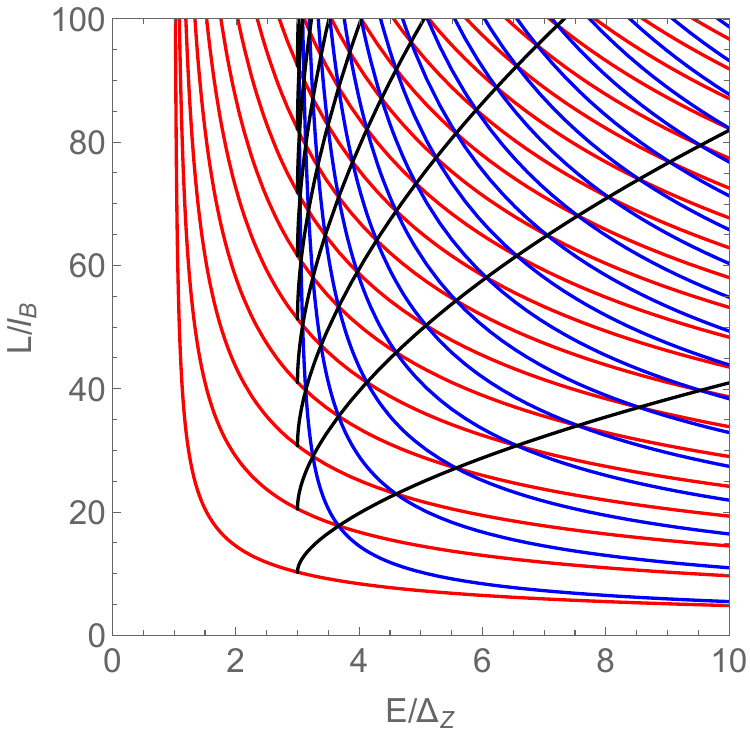}
	c)\includegraphics[width=0.32\textwidth]{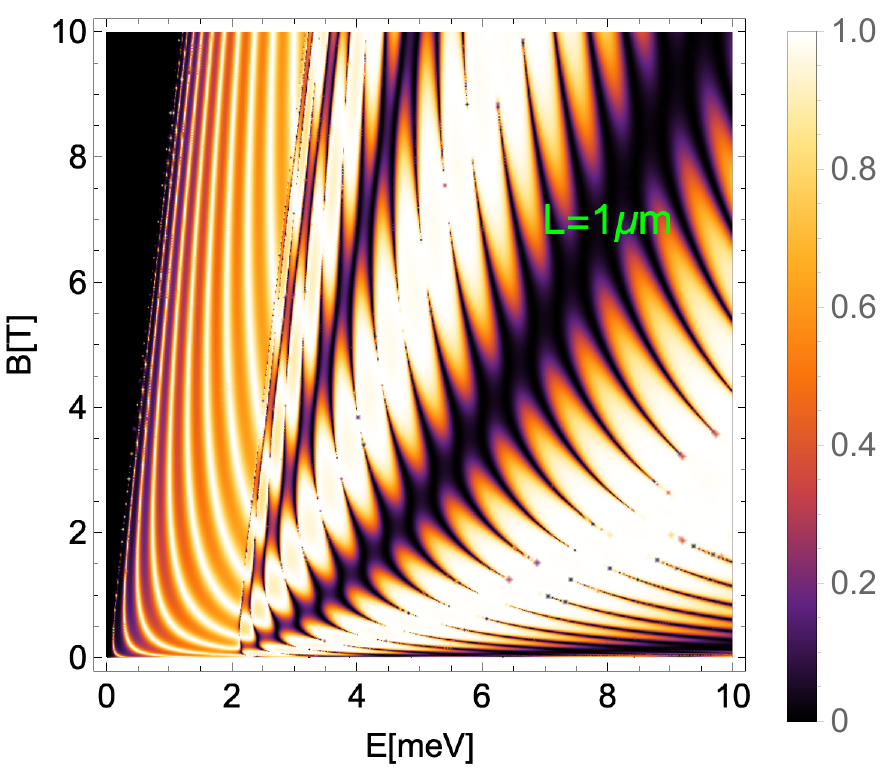}
	\caption{a) Transmission probabilities for either (i) a spin magnon in the  $\nu=-1$ and $\nu=0$ F phase (blue region in Fig. \ref{fig:LLSW}.a) or (ii) a pseudospin magnon in the $\nu=-1$ and $\nu=0$ CDW phase (orange region in Fig. \ref{fig:LLSW}.a) at normal incidence ($k_y=0$) and $B=10T$ as a function of the energy of the magnon and the length $L$ in units of the magnetic length $l_B=\sqrt{\frac{eB}{\hbar}}$. The symmetry breaking terms have been chosen such that $\Delta_0^1=\Delta_Z$ and $\Delta_0^2=3\Delta_Z$. b) Zeroes of $\sin(k_{x,0}^1L)$ (red) and $\sin(k_{x,0}^2L)$ (blue). When both vanish at the same time, this leads to the valleys and crests  (black) in the transmission described by Eq. (\ref{eq:maxima}). c) Transmission probability for a spin magnon in the F phase at normal incidence as a function of the magnetic field. The gaps are shifted with the magnetic field due to the Zeeman effect.}
	\label{fig:scat}
\end{figure*}

\begin{table}
	\centering
	\begin{tabular}{c|c|c|}
		\cline{2-3}
		& $\Delta_0^1$ & $\Delta_0^2$ \\ \hline
		\multicolumn{3}{|c|}{Spin modes} \\
		\hline
		\multicolumn{1}{ |c|  }{F} & $\Delta_Z$ & $\Delta_Z+2u_\perp$ \\
		\hline
		\multicolumn{1}{ |c|  }{CAF} & $0$ & $\Delta_Z$ \\
		\hline
		\multicolumn{3}{|c|}{Pseudospin modes} \\
		\hline
		\multicolumn{1}{ |c|  }{CDW} & $-u_\perp-u_z+\Delta_\text{AB}$ & $u_\perp-u_z+\Delta_\text{AB}$ \\
		\hline
		\multicolumn{1}{ |c|  }{KD}& $0$ & $\sqrt{2u_\perp(u_\perp+u_z)}$  \\
		\hline
	\end{tabular}
	\caption{Gaps $\Delta_\nu^\alpha$ at $\nu=0$ for $\alpha\in\{1,2\}$ corresponding to the two spin modes in the F and CAF phases and the two pseudo-spin modes in the CDW and CKD phases. The remaining modes ($\alpha\in\{3,4\}$) are entanglement modes.}
	\label{tab:gaps}
\end{table}

\section{Beating interference pattern}

We now apply boundary conditions which consist of matching the SU(4)-rotated spinorial wavefunctions (see Appendix \ref{app:matching}) and their derivatives at the interfaces. This procedure is valid as long as the interface is sharp on a length scale that is set by the magnon wavelength, $\lambda=2\pi/k$. This is indeed the case for low-energy magnons described by the dispersion (\ref{eq:disp}) where $\lambda\gg l_B$ is required. Notice that the interfaces are defined by gates that are separated from the graphene sheet by distances on the order of some ten nanometers. Furthermore, we neglect, here, a possible coupling between the magnon and the edge state at the interface, which may be dressed by a spin or pseudospin texture. While a microscopic investigation of such coupling, which is beyond the scope of the present paper, would in principle be of interest, one may expect that it is relatively weak. Indeed, we expect it to be governed by the small ratio $l_B/\lambda$ between the characteristic width of the edge state $\sim l_B$ and $\lambda$. 

Solving the resulting system of linear equations, we find that the only non-zero coefficients are $r_i$ and $t_i$ (the reflection and transmission of the spin or pseudospin magnon), while in the central region, the non-zero coefficients are $c_1^\pm$ and $c_2^\pm$, corresponding to the two spin or pseudospin modes at $\nu=0$. We obtain thus the main result of our paper, namely the expression for the numerator of the transmission amplitude of a two-mode magnonic Faby-P\'erot cavity valid when the $\nu=-1$ and $\nu=0$ regions are in the same phase (the full expression is given in Appendix \ref{app:matching})
\begin{align}
	t_i&\propto4ik_{x,-1}^i[k_{x,0}^1\sin(k_{x,0}^2L)+k_{x,0}^2\sin(k_{x,0}^1L)],
	\label{eq:trans}
\end{align}
where $i\in\{1,2\}$. Fig. \ref{fig:scat}.a) shows the transmission probability $|t_i|^2$ at normal incidence ($k_y=0$) for either (i) a spin magnon when both the lateral $\nu=-1$ and the central $\nu=0$ regions are in the F phase or (ii) a pseudospin magnon when all regions are in the CDW phase. Both cases are caracterized by two gaps associated with the two (i) spin or (ii) pseudospin modes (See Table \ref{tab:gaps}). We observe two thresholds corresponding to the two gaps. Below $\Delta_0^1$, which is identical to the spin or pseudospin gap $\Delta_{-1}^\mu$ at $\nu=-1$, there is a very small transmission probability for small length of the $\nu=0$ region which corresponds to evanescent modes. For longer lengths, there is no transmission since both modes are gapped. Between $\Delta_0^1$ and $\Delta_0^2$, only one mode is transmitted and we observe the usual Fabry-P\'erot interference pattern. However, above $\Delta_0^2$, both modes can propagate and we observe a beating pattern with valleys and crests in the transmission. This pattern can be understood simply by analyzing the zeros of Eq. (\ref{eq:trans}). It vanishes when both $\sin(k_{x,0}^1L)$ and $\sin(k_{x,0}^2L)$ are equal to zero. The red and blue curves in Fig. \ref{fig:scat}.b) correspond respectively to $L^1_m(E)=\frac{m\pi}{k_{x,0}^1(E)}$ and $L^2_n(E)=\frac{n\pi}{k_{x,0}^2(E)}$ with $m,n\in\mathbb{N}^*$. When both terms vanish, we observe either a valley in the transmission probability if the denominator of Eq. (\ref{eq:trans}) is non-zero, or a crest if the denominator vanishes too. The valleys and crests of the beating pattern are represented as the black curves in Fig. \ref{fig:scat}.b) and obey the equation (for $k_y=0$)
\begin{align}
	\frac{L_\text{max}(E)}{l_B}=\frac{m\pi\left(E-\Delta_0^1+\sqrt{(E-\Delta_0^1)(E-\Delta_0^2)}\right)}{k_{x,0}^1(E)(\Delta_0^2-\Delta_0^1)}.
	\label{eq:maxima}
\end{align}
In order to make a connection with a typical experimental situation, in which it is difficult to modify \textit{in situ} the junction length $L$, we plot in Fig. \ref{fig:scat}.c) the transmission probability of the junction for a spin magnon in the F phase as a function of the energy and the magnetic field for a $1\mu m$ long central region and a characteristic value $u_\perp=1$ meV. We observe a similar pattern as a function of the magnetic field since changing the magnetic field modifies the magnetic length while maintaining the length of the junction fixed. Since the gaps depend on the Zeeman term, we can see that they increase with the magnetic field. For a pseudospin magnon in the CDW phase, the interference pattern is similar except that the gaps in Tab. \ref{tab:gaps} depend on the magnetic field only through $u_\perp$ and $u_z$. Finally, we mention that the experimental observation of such an interference pattern allows one to obtain a quantitative measurement of the pseudospin anisotropic parameters. For example, in the F and the CDW phase, we have $\Delta_0^2-\Delta_0^1=2u_\perp$. In the KD phase, there is a gapless and linearly dispersing pseudospin magnon at $\nu=-1$ and $\nu=0$ with vanishing gap such that the transmission probability of a pseudospin magnon is identical to Fig. \ref{fig:scat}.a) with $\Delta_0^1=0$ such that the interference pattern appears at $E=0$ and the beating pattern appears at $\Delta_0^2$.

\section{Experimental relevance}

\begin{figure}[h]
	\begin{center}
		\includegraphics[width=8cm]{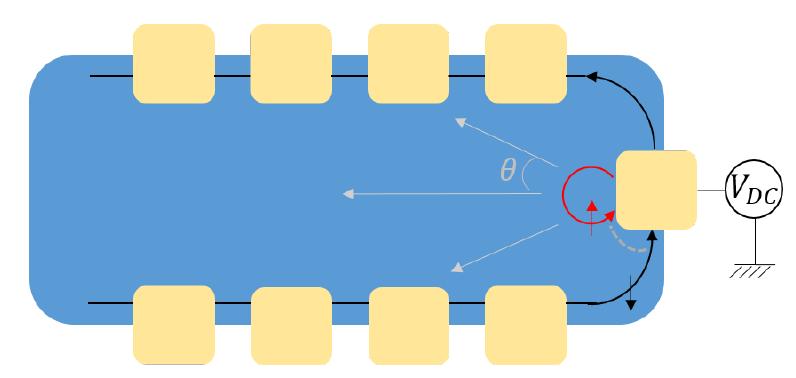}
		\caption{Schematic representation of a device that would generate normal incidence magnons}
		\label{fig:normal_incidence}
	\end{center}
\end{figure}

Magnon injection and detection in a graphene sheet can be achieved via local gates in the vicinity of which the filling factor corresponds to a completely filled or empty $n=0$ LL, \textit{i.e.} at $\nu=2$ or $\nu=-2$. In the present case, where we discuss quantum Hall ferromagnetism at $\nu=-1$, the latter local filling is of interest. A magnon can now be emitted via charge injection from the additional edge channel associated with this local filling factor and sketched by the red circle in Fig. \ref{fig:normal_incidence} to the adjacent (black) channel, which has an opposite spin or pseudospin orientation. The difference in spin between the charges in the edge channels is precisely compensated by the spin carried by the magnon. In order to allow for this injection to happen, the energy difference between the two channels $E=-eV_{DC}$ tuned by the gate voltage $V_{DC}$ must overcome the relevant gap $\Delta_\nu^\alpha$ for a magnon of type $\alpha$ in the region of filing $\nu$. Detection of magnons takes place at another contact via the same spin- or pseudospin-flip mechanism, and the electric charge signal associated with the electron transfer between the edge channels is proportional to the transmission coefficient $|t_i|^2$, which was calculated in the preceding section. 

Here, we have focused on normal incidence mainly for illustration reasons and because of its experimental relevance. Indeed the setup sketched in Fig. \ref{fig:normal_incidence} shows lateral ohmic contacts on the lower and upper edge. They may in principle allow one to filter magnons and thus to effectively collimate them to normal incidence.

\section{Summary}

In conclusion, we have shown that graphene is a promising platform for the emerging field of coherent SU(4) magnonics. Since one cannot polarize both spin and pseudospin simultaneously at $\nu=0$, disentangled ground states are either spin or pseudospin polarized and thus only support spin or pseudospin waves, respectively. An incoming magnon at $\nu=\pm1$ can thus propagate in the $\nu=0$ region only if (i) the ground state is identical on both sides and (ii) the magnon type (spin or pseudospin) is identical to the ground state polarization, which already gives valuable information about the ground state. Moreover, because there are always two (spin or pseudospin) modes with different gaps and wavevectors at $\nu=0$, we predict a beating interference pattern in a Fabry-P\'erot cavity. Its observation would allow for an experimental determination of the anisotropic parameters $u_\perp$, $u_z$ and $\Delta_\text{AB}$. Transmission in the case of a KD phase is special in that there exists a gapless mode such that the transmission is non-zero down to $E=0$, but one retrieves the characteristic beating pattern above the second gap $\Delta_0^2$.

\section{Acknowledgments}

We acknowledge financial support from Agence Nationale de la Recherche (ANR project ``GraphSkyrm'') under Grant No.  ANR-17-CE30-0029. We thank Fran\c cois Parmentier for fruitful discussions and his valuable scientific input. Finally, we thank Patrice Jacques for his assistance with our figures.

\appendix

\section{Non-linear sigma model and phase diagrams}

\label{app:nlsm}

The order parameter of the quantum Hall ferromagnet is the $4\times 4$ matrix field
\begin{align}
	P(\mathbf{r},t)=Z(\mathbf{r},t)Z^\dagger(\mathbf{r},t)
\end{align}
which is a  a projector that obeys $P^2=P$, $P^\dagger=P$, while $Z(\mathbf{r},t)$ is a $4\times\tilde{\nu}$ matrix where we have defined $\tilde{\nu}=2+\nu$ which counts the number of filled sub-LLs starting from the empty $N=0$ LL.

Following Kharitonov\cite{Kharitonov2012}, we consider the non-linear sigma model Lagrangian which is composed of three terms, the the Berry phase term $\mathcal{L}_\text{BP}$, the SU(4) invariant non-linear sigma model term $\mathcal{L}_\text{NLSM}$ and the low-energy anisotropies $\mathcal{L}_\text{A}$ such that the total Lagrangian of the system is
\begin{align}
	\mathcal{L}=\mathcal{L}_\text{BP}+\mathcal{L}_\text{NLSM}+\mathcal{L}_\text{A}
	\label{eq:NLSM}
\end{align}
with
\begin{widetext}
	\begin{subequations}
		\begin{align}
			\mathcal{L}_\text{BP}&=Sn_0\int d^2rZ^\dagger i\partial_tZ, \\
			\mathcal{L}_\text{NLSM}&=\rho_s\int d^2r\text{Tr}\left[\bm{\nabla}P\bm{\nabla}P\right]=2\rho_s\int d^2r\partial_jZ^\dagger(1-ZZ^\dagger)\partial_jZ \\
			\mathcal{L}_\text{A}&=n_0\int d^2r\left\{\frac{1}{2}\sum_{a=x,y,z}u_at_a(P)-\Delta_Z\text{Tr}[\sigma_zP]-\Delta_\text{AB}\text{Tr}[\tau_zP]\right\}
		\end{align}
		\label{eq:Lagrangian}
	\end{subequations}
\end{widetext}
where $S=\frac{1}{2}$ is the spin of the electron, $n_0=(2\pi l_B^2)^{-1}$ is the electronic density in a sub-LL, $\rho_s=\frac{1}{16\sqrt{2\pi}}\frac{e^2}{\varepsilon l_B}$ is the spin stiffness, $\Delta_Z=g\mu_BB$ is the Zeeman term, $\Delta_\text{AB}$ is an on-site potential with opposite sign on the A and B sublattices (e.g. for graphene on hBN) and due to to the sublattice-valley correspondence, it is represented by the Pauli matrix $\tau_z$ which acts in valley space. $u_x=u_y\equiv u_\perp$ and $u_z$ are the valley anisotropies\cite{Kharitonov2012} with
\begin{align}
	t_a(P)=\text{Tr}[\tau_aP]^2-\text{Tr}[(\tau_aP)^2]
\end{align}
We also label the ground state order parameter (in the absence of excitations) as $P_0$ such that the phase diagrams presented in the next sections are computed with $P_0$.

\subsection{Filling factor $\nu=-1$}

At filling factor $\nu=-1$, because there is only one sub-LL that is filled, the symmetry breaking mechanism is
\begin{align}
	SU(4)\rightarrow SU(3)\otimes U(1)
\end{align}
where $SU(3)$ represents the invariance of the ground state under rotations between the filled levels and $U(1)$ the global phase of the empty level. The field $Z$ is thus an element of the complex projective space $CP^3=SU(4)/SU(3)\otimes U(1)$ which has dimension $d=6$. Because only one level is filled, the field $Z(\mathbf{r},t)$ is a four-component spinor. In the ground state, we have $Z(\mathbf{r},t)=|F\rangle$ as can be seen from Fig. \ref{fig:LLSW}.b) of the main text.

In order to describe the spinor $F$, which must be parametrized by $d=6$ parameters, we express it as a Schmidt decomposition in the basis $\{|K\uparrow\rangle,|K\downarrow\rangle,|K'\uparrow\rangle,|K'\downarrow\rangle\}$ as\cite{Doucot2008,Lian2017,Atteia2021a}
\begin{align}
	|F\rangle=\cos\frac{\alpha}{2}|\mathbf{n}\rangle|\mathbf{s}\rangle+e^{i\beta}\sin\frac{\alpha}{2}|-\mathbf{n}\rangle|-\mathbf{s}\rangle,
	\label{eq:param}
\end{align}
where $|\mathbf{n}\rangle|\mathbf{s}\rangle=|\mathbf{n}\rangle\otimes|\mathbf{s}\rangle$ is the tensor product of the spinors 
\begin{align}
	|\mathbf{n}\rangle&=\begin{pmatrix}
		\cos\frac{\theta_P}{2} \\ \sin\frac{\theta_P}{2}e^{i\varphi_P}
	\end{pmatrix}, 
	\label{eq:param1} \\
	|\mathbf{s}\rangle&=\begin{pmatrix}
		\cos\frac{\theta_S}{2} \\ \sin\frac{\theta_S}{2}e^{i\varphi_S}
	\end{pmatrix}, \label{eq:param2} 
\end{align}
acting in valley and spin spaces respectively with $\theta_{S,P}$ and $\varphi_{S,P}$ the SU(2) parameters of the spin and valley Bloch spheres respectively, while $\alpha$ and $\beta$ are dubbed the "entanglement" parameters (which are non-zero only in the CAF and AFI phases).

\begin{figure}[h!]
	\begin{center}
		a)\includegraphics[width=0.35\textwidth]{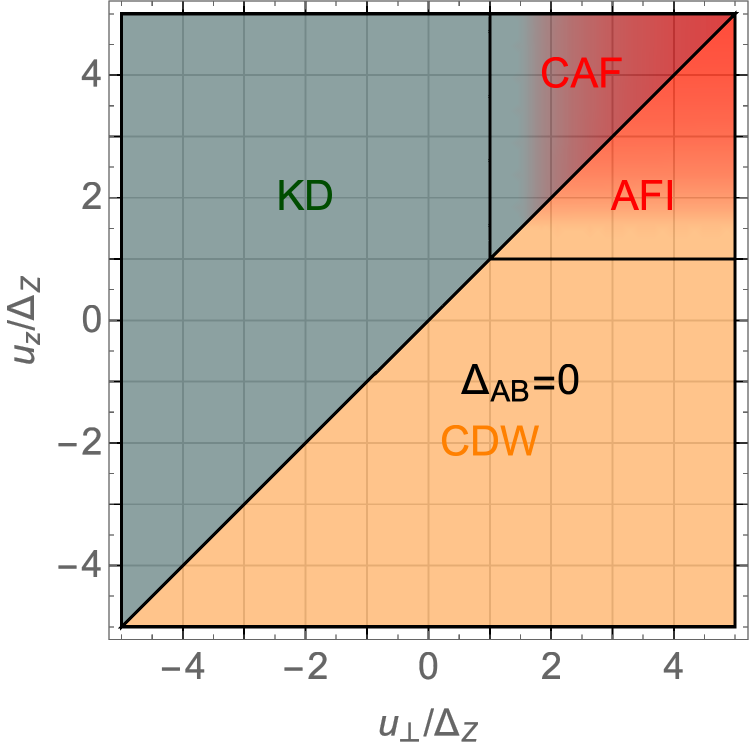}
		a)\includegraphics[width=0.35\textwidth]{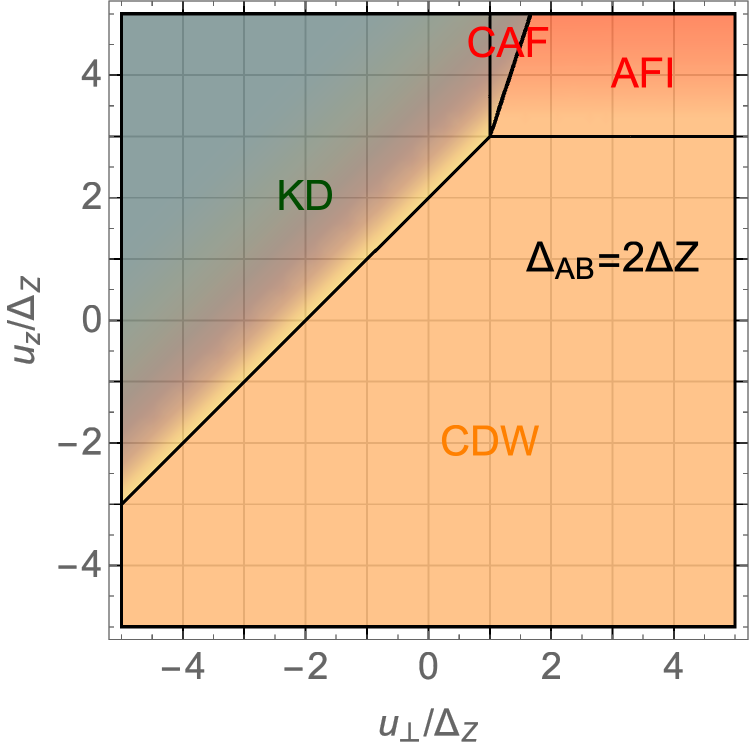}
		\caption{Phase diagram at $\nu=-1$ as a function of $u_\perp$ and $u_z$ for a) $\Delta_\text{AB}=0$ and b) $\Delta_\text{AB}=2\Delta_Z$}
		\label{fig:Diag_nu1}
	\end{center}
\end{figure}

Fig. \ref{fig:Diag_nu1} shows the ground state phase diagram at $\nu=-1$ in the absence and the presence of the sublattice potential $\Delta_{AB}$. It is composed of four phases : (i) two ferromagnetic phases, the charge density wave (CDW) and the Kékule distortion phases, and (ii) two entangled phases, the canted anti-ferromagnetic phase (CAF) and the anti-ferrimagnetic phase (AFI). In the absence of a sublattice potential, the transition between the ferromagnetic phases is of first order while the transition between the ferromagnetic and entangled phases are of second order due to the competition between the Zeeman term and the anistropic parameters $u_\perp$ and $u_z$.

\subsection{Filling factor $\nu=0$}

At filling factor $\nu=0$, two sub-LL are occupied such that the symmetry breaking mechanism is
\begin{align}
	SU(4)\rightarrow SU(2)\otimes SU(2)\otimes U(1)
\end{align}
where the two $SU(2)$ groups represent the invariance of the order parameter under rotations among the filled and the empty states, while $U(1)$ represents their relative phase. The coset space made of the generators of the broken symmetries is the Grassmanian $\text{Gr}(2,4)=U(4)/U(2)\otimes U(2)$ which has dimension 8. Because half of  the generators are canonically conjugate to the other half, there are thus 4 magnon modes. An element of the Grassmanian $\text{Gr}(2,4)$ is a $4\times 2$ matrix
\begin{align}
	Z=(Z_1,Z_2)=\begin{pmatrix}
		Z_{11}&Z_{12}\\
		Z_{21}&Z_{22}\\
		Z_{31}&Z_{32} \\
		Z_{41}&Z_{42}
	\end{pmatrix}=(Z_{\alpha n})
\end{align}
where $Z_1$ and $Z_2$ are two orthogonal spinors which describe the two filled sub-LL. An important property of the Grassmanian is that the electrons in the filled levels are indistinguishable, namely, the order parameter $P$ is invariant under unitary transformations that mix the two levels
\begin{align}
	Z'=ZU\quad\Longrightarrow \quad P'=Z'Z'^\dagger=P
	\label{eq:indistinguishable}
\end{align}
where $U$ is a $2\times 2$ unitary matrix.

In the ground state, two sub-LL are filled according to the orthogonal spinors $|F_1\rangle$ and $|F_2\rangle$. We parametrize these spinor with 8 angles according to the dimension of the Grassmanian
\begin{align}
	|F_1\rangle&=\cos\frac{\alpha_1}{2}|\mathbf{n}\rangle|\mathbf{s}\rangle+e^{i\beta_1}\sin\frac{\alpha_1}{2}|-\mathbf{n}\rangle|-\mathbf{s}\rangle \label{eq:Schmidt1} \\
	|F_2\rangle&=\cos\frac{\alpha_2}{2}|\mathbf{n}\rangle|-\mathbf{s}\rangle+e^{i\beta_2}\sin\frac{\alpha_2}{2}|-\mathbf{n}\rangle|\mathbf{s}\rangle.
	\label{eq:Schmidt2}
\end{align}

\begin{figure}[b]
	\begin{center}
		b)\includegraphics[width=0.35\textwidth]{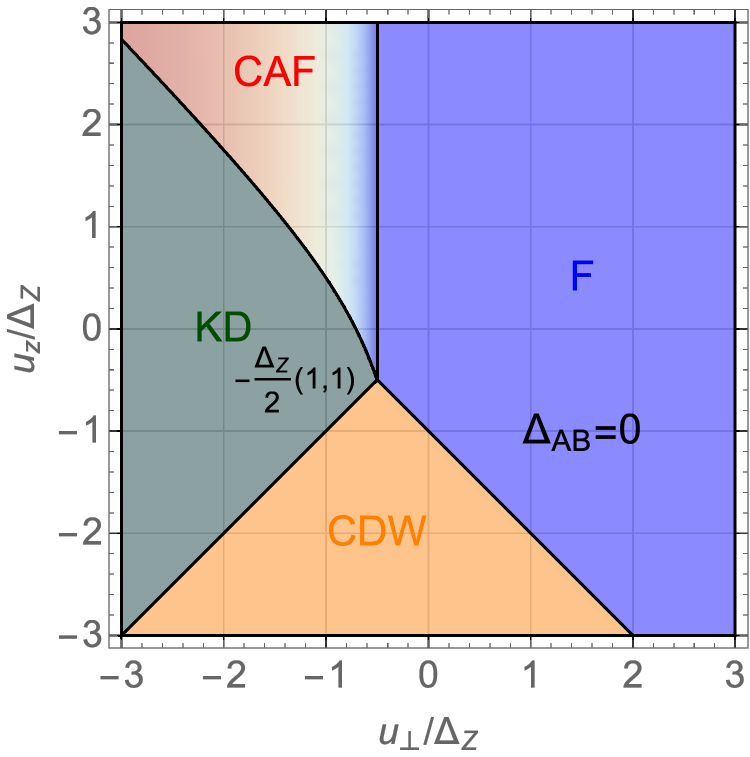}  
		b)\includegraphics[width=0.35\textwidth]{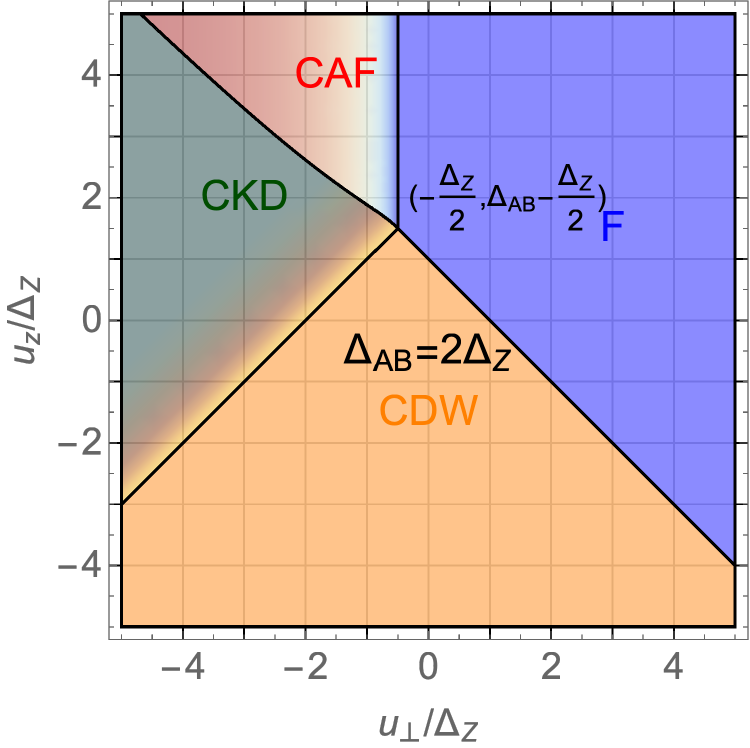}  
		\caption{Phase diagram at $\nu=0$  for a) $\Delta_\text{AB}=0$ and b) $\Delta_\text{AB}=2\Delta_Z$}
		\label{fig:Diag_nu0}
	\end{center}
\end{figure}

The order parameter of the ground state is the projector 
\begin{align}
	P_0=Z^\dagger Z=|F_1 \rangle\langle F_1|+ |F_2\rangle \langle F_2|
\end{align}
Fig. \ref{fig:Diag_nu0} shows the ground state phase diagram at $\nu=0$ in the absence and presence of the sublattice potential $\Delta_{AB}$. In the absence of a sublattice potential, all transitions are of first order except for the F|CAF transition. When $\Delta_{AB}$ is introduced, we have a second order transition between the CDW and CKD (canted Kékule distortion) phases analogously to the $\nu=-1$ case.

\section{Matching at the interfaces}

\label{app:matching}

The boundary conditions are obtained by matching the SU(4) rotation operators and their derivatives at the interfaces at the interfaces, namely
\begin{align}
	\Psi^L(x=0,y,t)\cdot \bm{\Gamma}_{-1}&=\Psi^C(x=0,y,t)\cdot\bm{\Gamma}_0 \\
	\Psi^C(x=L,y,t)\cdot\bm{\Gamma}_0&=\Psi^L(x=L,y,t)\cdot \bm{\Gamma}_{-1} \\
	\partial_x\Psi^L(x=0,y,t)\cdot \bm{\Gamma}_{-1}&=\partial_x\Psi^C(x=0,y,t)\cdot\bm{\Gamma}_0 \\
	\partial_x\Psi^C(x=L,y,t)\cdot\bm{\Gamma}_0&=\partial_x\Psi^L(x=L,y,t)\cdot \bm{\Gamma}_{-1}
\end{align}
with $\bm{\Gamma}_{-1}=(\Gamma_1,\Gamma_2,\Gamma_3,\Gamma_1^\dagger,\Gamma_2^\dagger,\Gamma_3^\dagger)$ and $\bm{\Gamma}_{0}=(\Gamma_{11},\Gamma_{12},\Gamma_{21},\Gamma_{22},\Gamma_{11}^\dagger,\Gamma_{12}^\dagger,\Gamma_{21}^\dagger,\Gamma_{22}^\dagger)$. Finally, solving this system of equation, we find that the only non-vanishing transmission coefficient in the right region corresponds to the mode $i$ with ($i=1$) for a spin magnon and ($i=2$) for a pseudospin magnon

\begin{widetext}
	\begin{subequations}
		\begin{align}
		t_i&=\frac{4ik_{x,-1}^i(k_{x,0}^1\sin(k_{x,0}^2L)+k_{x,0}^2\sin(k_{x,0}^1L)}{A+iB} \\
	A&=[(k_{x,0}^1)^2+(k_{x,0}^2)^2+4(k_1^i)^2]\sin(k_{x,0}^1L)\sin(k_{x,0}^2L)+2k_{x,0}^1k_{x,0}^2[1-\cos(k_{x,0}^1L)\cos(k_{x,0}^2L)] \\
	B&=4k_1^i[k_{x,0}^1\cos(k_{x,0}^1L)\sin(k_{x,0}^1L)+k_{x,0}^2\cos(k_{x,0}^2L)\sin(k_{x,0}^1L)]
		\end{align}
	\end{subequations}
\end{widetext}

\bibliographystyle{apsrev4-1}
\bibliography{library}

\end{document}